\documentclass[prc,twocolumn,showpacs,preprintnumbers,amsmath,amssymb,superscriptaddress,floatfix,nofootinbib]{revtex4}
\usepackage{graphicx}
\usepackage{amsmath}
\usepackage{amsfonts}
\usepackage{slashed}
\usepackage{amssymb}
\usepackage{hyperref}
\newcommand{\smsm}[1]{\scriptscriptstyle{\scriptscriptstyle{#1}}}

\newcommand{\boldm}[1]{\mbox{\boldmath{$#1$}}}
  \def\CL{{\cal L}}
\def\CM{{\cal M}}

\begin{document}

\title{Triangle singularity in the $J/\psi \to \gamma \bar{p} \Delta$ decay}

\author{Ke Wang}\email{wangke563@qq.com}\author{Rong Li}\email{rongliphy@xjtu.edu.cn}\author{Bo-Chao Liu} \email{liubc@xjtu.edu.cn}
\affiliation{MOE Key Laboratory for Nonequilibrium Synthesis and Modulation of Condensed Matter, School
of Physics, Xi’an Jiaotong University, Xi ’an 710049, China.}
\affiliation{
Institute of Theoretical Physics, Xi ’an Jiaotong University, Xi ’an 710049, China.
}

\begin{abstract}
In this work, we study the role of triangle singularity in the $J/\psi \to \gamma \bar{p} \Delta$ decay. We find that through a triangle mechanism, involving a triangle loop composed by $\omega$, $\pi$ and $p$, this decay may develop a triangle singularity and produce a visible peak in the invariant mass $M_{\gamma\Delta}$ around 1.73 GeV with a width of 0.02 GeV. Such a triangle mechanism may also cause significant spin effects on the final $\Delta$, which can be detected by measuring its spin density matrix elements. Our calculations show that the branching ratios due to the triangle mechanism is  Br($J/\psi\to \gamma \bar p\Delta,\Delta\to \pi p$)=$1.058\times 10^{-6}$. Hopefully, this reaction can be investigated at BESIII and future experiments, e.g. Super Tau-Charm Facility, and the narrow width of the induced structure, the moving TS position and the distinct features of the spin density matrix elements of the $\Delta$ may serve as signals for the triangle singularity mechanism. 
\end{abstract}

\maketitle

\section{INTRODUCTION}
\label{introductiuon}

Triangle singularity(TS) as one kind of kinematical singularities in the scattering amplitude was first studied by Landau in 1956\cite{Landau:1959fi}. Later, the corresponding physical picture of the special kinematic conditions needed to produce TS, known as the Coleman-Norton theorem, was described in Ref.\cite{Coleman:1965xm}. Specifically, for the decay process $A \to B+C$ proceeding through a triangle loop composed by internal particles 1, 2 and 3, the particle $A$ first decays into particles 1 and 2, then particle 1 decays into the particle 3 and $B$, finally the particle 2 and 3 merge into the particle $C$. TS occurs in the amplitude only when these sub-processes take place in a classical manner. It corresponds to the case that all three intermediate particles are on shell simultaneously and their three momenta are collinear in the rest frame of particle $A$. Besides, particle 3 must move fast enough to catch up with particle 2 and merge into particle $C$.

In recent years, TS has attracted a lot of attentions of researchers and has been suggested to play an essential role for understanding the nature of some observed structures and clarifying some important puzzles\cite{Wu:2011yx,Aceti:2012dj,Wu:2012pg,Achasov:2015uua,Du:2019idk,Liang:2019yir,Jing:2019cbw,Guo:2019qcn,Sakai:2020ucu,Molina:2020kyu,Sakai:2020crh,Yan:2022eiy,Wang:2013cya,Liu:2013vfa,Nakamura:2019btl,Liu:2020orv,Braaten:2022elw,Achasov:2022onn}. For example, the abnormally large isospin-breaking effects observed in $J/\psi \to \gamma \eta(1405) \to \gamma\pi^0 f_0(980)$ can be understood by considering the TS mechanism originating from the $K^*\bar{K}K$ loop\cite{Wu:2011yx,Aceti:2012dj,Wu:2012pg,Achasov:2015uua,Du:2019idk,Liang:2019yir}. The band around 1.4 GeV on the $\pi^0\phi$ distribution in Dalitz plot for the isospin-breaking decay $J/\psi \to \eta\pi^0\phi$ can also be explained by the TS mechanism\cite{Jing:2019cbw}. Furthermore, some exotic states observed recently in experiments, e.g. $Z_c$\cite{Wang:2013cya,Liu:2013vfa,Nakamura:2019btl}, $X(2900)$\cite{Liu:2020orv} and $T^+_{cc}$\cite{Braaten:2022elw,Achasov:2022onn}, have been argued to involve TS mechanism. For a comprehensive review of these topics, we refer to Ref.\cite{Guo:2019twa}.

Although TS mechanism may be essential for understanding those interesting and important experimental phenomena, further studies are still needed to investigate its physical effects and find ways to identify its contribution in experiments. It is well known that TS mechanism can cause an enhancement in the invariant mass spectrum of final particles, which has been the main focus of previous studies\cite{Liu:2015taa,Huang:2021olv,Szczepaniak:2015hya,Guo:2016bkl,Wang:2016dtb,Xie:2016lvs,Liang:2017ijf,Pavao:2017kcr,Roca:2017bvy,Debastiani:2017dlz,Xie:2017mbe,Bayar:2017svj,Sakai:2017hpg,Dai:2018hqb,Liang:2019jtr,Nakamura:2019emd,Liu:2019dqc,Sakai:2020fjh,Shen:2020gpw,Huang:2020kxf,Luo:2021hyy}. However, since TS and resonances can induce similar structures in the invariant mass spectrum, it raises the question on how to distinguish these two mechanisms. One possible way is to change the kinematic conditions that are necessary for TS mechanism\cite{Guo:2019twa,Jing:2019cbw,Liu:2015taa,Huang:2021olv}. The structure should disappear for the TS model but not for the resonance model when changing the kinematic conditions. Although this method is feasible in principle, it changes the conditions of the original experiment and may introduce other ambiguities, e.g., the change of relative strength of various contributions due to varying kinematic conditions. Therefore, a better method would be one that can distinguish these two mechanisms without changing the experiment conditions. In our recent work\cite{Wang:2022wdm}, we suggest that in some cases TS mechanism may cause significant spin effects, which offers an alternative way to verify TS mechanism and thus deserves further studies. 

In this work, we propose that in the radiative decay process $J/\psi \to \gamma \bar{p} \Delta(1232)$ the TS mechanism, through the triangle loop involving $\omega$, $\pi$ and $p$ as shown in Fig.\ref{Feynman3}, may play an important role. In this process, the couplings of the three vertices $J/\psi \to p\bar{p}\omega$, $\omega\to \gamma\pi$ and $\pi p \to \Delta$(denoting the $\Delta(1232)$ hereafter) involved in the loop are relatively strong\cite{ParticleDataGroup:2022pth}. Furthermore, the small width of the intermediate states in the loop may also enhance the triangle loop contribution and can produce a relatively narrow peak in the $\gamma \Delta$ invariant mass spectrum at the position of the TS. At the same time, as argued in our previous work such a TS mechanism may also cause significant spin effects. The physical picture behind this expectation is quite simple. When incident particles are moving along some fixed direction, the produced intermediate state may have spin alignment due to angular moment conservation. For example, considering the $\Delta$ resonance produced in $\pi N$ elastic scattering process in the center of mass frame, the spin projection on the z-axis of the produced $\Delta$ can only be $\pm\frac{1}{2}$ if we take z-axis along the beam direction. Therefore, the spin of the $\Delta$ is aligned and the angular distribution of its decay products is anisotropic. The spin status of the $\Delta$ can be described by the spin density matrix elements(SDMEs) and measured through the analysis of the angular distribution of the $\Delta\to \pi N$ decay. In the $J/\psi \to \gamma \bar p \Delta$ process through the triangle diagram, according to the Coleman-Norton theorem, the $\pi$ and $N$ in the loop should move along the direction of the momentum of the $\Delta$ at TS in the $\gamma\Delta$ rest frame. It means, if we consider the helicity states of the $\Delta$, i.e. choosing the quantization axis along the direction of the momentum of the $\Delta$, the helicity should be $\pm\frac{1}{2}$ similar as the case of the $\Delta$ production in the $\pi N$ elastic scattering process mentioned above. In other words, the special kinematic conditions required by the TS constrain the helicity of the $\Delta$ in the $\gamma \Delta$ rest frame, which is absent for other mechanisms. Therefore, if the TS mechanism indeed plays an important role in this reaction, we expect a peak structure in the $\gamma \Delta$ invariant mass spectrum and the production of the $\Delta$ with helicity $\pm\frac{1}{2}$ should be enhanced near TS.

This paper is organized as follows. In Sec.\ref{model}, we present the theoretical framework and amplitudes for the reaction $J/\psi \to \gamma \bar{p} \Delta$. In Sec.\ref{results}, we show the numerical results and discuss their implications. Finally, we summarize our findings and conclusions in Sec.\ref{summary}.

\section{MODEL AND INGREDIENTS}
\label{model}

\begin{figure}[htbp] \begin{center} \includegraphics[scale=0.5]{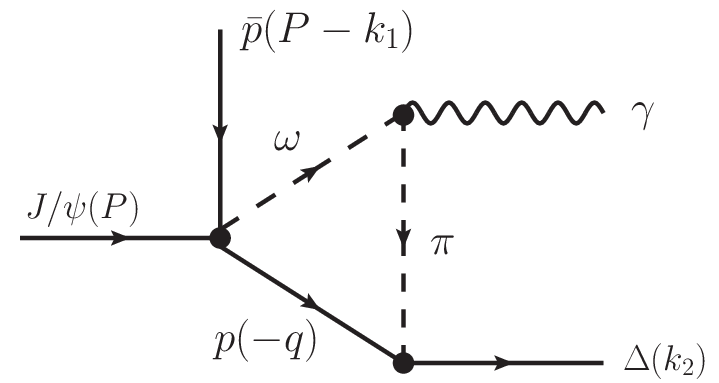} \caption{The Feynman diagram for the radiative decay process $J/\psi \to \gamma \bar{p} \Delta$ through a triangle loop involving the $\omega$, $\pi$ and $p$.} \label{Feynman3} \end{center} \end{figure}

In this work, we shall introduce the TS mechanism in the radiative decay process $J/\psi \to \gamma \bar{p} \Delta$ within an effective Lagrangian approach. The Feynman diagram for the process that may produce TS is shown in Fig.\ref{Feynman3}. In this process, the $J/\psi$ first decays into $p\bar{p}\omega$, then $\omega$ decays to a photon and a $\pi$ meson. In the $\gamma\Delta$ rest frame, if the $\pi$ meson travels along the momentum of the proton produced in $J/\psi$ decay and moves faster than it, the $\pi$ may catch up with the proton and they can finally merge into the final $\Delta$. According to the results in Ref.\cite{Bayar:2016ftu}, TS exists in this decay process only when the special kinematic conditions are satisfied. Using the method in Ref.\cite{Bayar:2016ftu}, if we adopt the nominal masses in PDG\cite{ParticleDataGroup:2022pth} for the involved particles in Fig.\ref{Feynman3}, it turns out that the TS should occur at $M_{\gamma\Delta}=1.731$GeV.

To calculate the decay amplitude for the Feynman diagram in Fig.\ref{Feynman3}, we need the Lagrangian densities for the various vertices. For the $J/\psi \to p\bar{p}\omega$ vertex, we adopt a contact interaction \begin{equation} \CL_{\psi\omega N\bar{N}} = g_c \bar{N} \psi^\mu \omega_\mu N, \end{equation} where $g_c$ is the coupling constant and can be determined through the $J/\psi \to p\bar{p}\omega$ partial decay width in PDG\cite{ParticleDataGroup:2022pth}. Note that up to now there is no evidence that resonance productions play an important role in the $J/\psi \to p \bar{p} \omega$ decay. For the $\omega\gamma\pi$ and $\Delta\pi N$ vertices, we adopt the effective Lagrangians\cite{Lu:2014yba,Zhao:2019syt,Fan:2019lwc,Xie:2014zga}. 
\begin{eqnarray}
    \CL_{\omega \gamma \pi} &=& \frac{e g_{\omega \gamma \pi}}{m_\omega} \varepsilon^{\mu \nu \alpha \beta} \partial_\mu \omega_\nu \partial_\alpha A_\beta \pi, \\
    \CL_{\Delta\pi N} & = & \frac{g_{\Delta\pi N}}{m_{\pi}}\bar{\Delta}^{\mu} \left(\vec{\tau} \cdot \partial_{\mu}\vec{\pi}\right) N+\text{h.c.},
\end{eqnarray}
where $A$ represents the photon field and $e$ is taken as $\sqrt{4\pi/137}$. The coupling constants $g_{\omega \gamma \pi}$ and $g_{\Delta\pi N}$ appearing in the above Lagrangian densities can be determined through the corresponding partial decay width using
\begin{eqnarray}
    \Gamma_{\omega \to \pi \gamma} &=& \frac{e^2 g^2_{\omega \gamma \pi}} {12\pi} \frac{|\boldm{p}_{\pi}|^3}{m^2_{\omega}}, \\
    \Gamma_{\Delta \rightarrow \pi N} & = & \frac{g_{\Delta\pi N}^{2}} {12 \pi} \frac{E_{N}+m_{N}}{m_\Delta m_{\pi}^{2}}|\boldm{p}_{\pi}|^3, \label{Gamma:Delta}
\end{eqnarray}
where $|\boldm{p}_{\pi}|$ and $E_N$ denote the magnitude of the three momentum of the $\pi$ and the nucleon energy in the rest frame of the mother particles, respectively. The obtained coupling constants are listed in Table \ref{tab1}.

\begin{table}[htbp]
    \caption{Coupling constants used in this work. The experimental decay widths are taken from Ref.\cite{ParticleDataGroup:2022pth}.}
    \begin{tabular}{ccccc}
        \hline\hline
        State   & Width               &  Decay      &      Adopted     &    $g$  \\
          & (MeV)               & channel     & branching ratio  &             \\
        \hline
        $ J/\psi$ & $9.26\times10^{-2}$ & $p\bar{p}\omega$ & $9.80\times10^{-4}$ & $7.30\times10^{-2}$   \\
        $ \omega$ &  8.68               & $\pi^0\gamma$    & $8.35\times10^{-2}$ & 1.83   \\
        $ \Delta$ &  117                & $N\pi$           & 0.994              & 2.07   \\
        \hline\hline
    \end{tabular}
    \label{tab1}
\end{table}

With the above Lagrangian densities for various vertices, we can straightforwardly obtain the amplitude for the triangle loop diagram in Fig.\ref{Feynman3} as
\begin{widetext}
    \begin{eqnarray}\label{Amplitude3}
    \CM^{T} &=&
        -i\frac{e g_c g_{\omega \gamma \pi} g_{\Delta\pi N}} {m_\pi m_\omega} \bar{u}^\mu_\Delta \varepsilon^\nu_{\psi} \varepsilon^{*\alpha}_{\gamma}\int\frac{\text{d}^4q}{(2\pi)^4} p_{\pi,\mu} G^{\frac{1}{2}}(p_p) \epsilon_{\beta\rho\lambda\alpha} p^\beta_\omega G^{1,\rho}_\nu(p_\omega) p^\lambda_\gamma G^0(p_\pi) F(p_\pi) v_{\bar{p}} \nonumber \\ & \equiv & g \bar{u}^\mu_\Delta \varepsilon^\nu_{\psi} \varepsilon^{*\alpha}_\gamma \CM_{\mu\nu\alpha} v_{\bar{p}},
    \end{eqnarray}
\end{widetext}
where $u_\Delta$,$v_{\bar p}$, $\varepsilon_\gamma$ and $\varepsilon_\psi$ are the spin functions of the $\Delta$, $\bar p$, photon and $J/\psi$, respectively. $G^J$s denote the propagators of the intermediate particles with spin $J$, which are defined as\cite{Fan:2019lwc,Chen:2020szc,Xie:2013db} 
\begin{eqnarray}
    G^0(q) &=& \frac{i}{q^2-m^2}\label{pp1},  \\
    G^1_{\mu\nu}(q) &=& -\frac{i(g_{\mu\nu}-q_\mu q_\nu/m^2)}{q^2-m^2}\label{pp2}, \\
    G^{\frac{1}{2}}(q) &=& \frac{i(\slashed{q}+m)}{q^2-m^2}\label{pp3},  
\end{eqnarray}
where $q$ and $m$ are the four momentum and the mass of the intermediate state. In the above amplitude, we have introduced a monopole form factor $F(p_\pi)$ for the intermediate $\pi$ meson in order to make the loop integral convergent, which is taken as\cite{Liu:2017vij,Liu:2017acq,Chen:2020zzz}
\begin{equation}
    F(p_\pi) = \frac{m^2_\pi-\Lambda^2_\pi}{p^2_\pi-\Lambda^2_\pi}.
\end{equation}
Here we note that near TS the off-shell effects of the intermediate states in the loop are small, so we do not need to consider the form factors for other particles. Furthermore, the possible problem of an artificial pole introduced by the form factor should not be worried here as discussed in Ref.\cite{Du:2019idk}. The cutoff $\Lambda_\pi$ can be determined through an empirical formula $\Lambda_\pi=m_\pi+\alpha \Lambda_{\text{QCD}}$\cite{Xiao:2018kfx,Huang:2020kxf,Ling:2021lmq}, where $\alpha$ is a dimensionless free parameter and $\Lambda_{\text{QCD}} =0.22$ GeV is the scale parameter of QCD. The $\alpha$ is usually taken to be about unity, and in this work we take $\alpha=1$ in the calculations.

\begin{figure}[htbp]
    \begin{center}
        \includegraphics[scale=0.5]{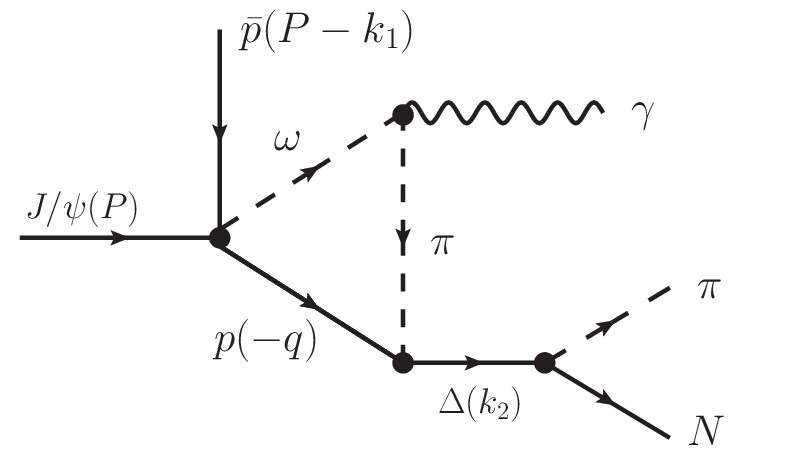}
        \caption{The Feynman diagram for the radiative decay process $J/\psi \to \gamma \bar{p} \pi N$ involving a triangle loop.}
        \label{Feynman4}
    \end{center}
\end{figure}

For the quasi three-body decay process, i.e. ignoring the decay of the $\Delta$, the invariant mass distribution of the $\gamma\Delta$ system can be obtained through the following formula\cite{ParticleDataGroup:2022pth,Jing:2019cbw}
\begin{eqnarray}\label{Gamma3}
    \frac{\text{d}\Gamma}{\text{d} M_{\gamma\Delta}} \! = \! \frac{4m_N m_\Delta}{(2 \pi)^{5} 2^4 m^{2}_{\psi}} \frac{\left|\boldm{p}_{\bar{p}}\right| \! \left|\boldm{p}_{\gamma}^{*}\right|}{3} \! \int \! \text{d} \Omega_{\bar{p}} \text{d} \Omega_{\gamma}^{*} \sum_{\text{spin}} \left|\CM^{T}\right|^{2} \! ,
\end{eqnarray}
where the quantities with or without $*$ represent that they are defined in the center of mass frame of the $\gamma\Delta$ system or the $J/\psi$ rest frame, respectively. To further consider the influences of the finite width effects of the $\Delta$ due to the $\Delta$ decay as shown in Fig.\ref{Feynman4}, we follow the approach used in Ref.\cite{Pavao:2017kcr} by introducing a mass distribution function for the $\Delta$ in Eq.\eqref{Gamma3}. Then we obtain

\begin{eqnarray}
    \frac{\text{d}\Gamma}{\text{d} M_{\gamma\pi N}} &=& \int \frac{4m_N M_{\pi N}}{(2 \pi)^{5} 2^4 m^{2}_{\psi}} \text{d} \Omega_{\bar p} \text{d} \Omega_\gamma^* \text{d}M_{\pi N}^2 \frac{|\mathbf{p}_{\bar p}| |\mathbf{p}_\gamma^*|}{3\pi} \nonumber \\ &&\times\frac{   m_\Delta \Gamma_\Delta\cdot \sum\limits_{\text{spin}}| \CM^{T}|^{2}}{(M_{\pi N}^2 - m^2_\Delta)^2 + (m_\Delta \Gamma_\Delta)^2},\label{4body}
    \end{eqnarray}
where $M_{\pi N}$ stands for the invariant mass of its decay products $\pi N$ or the varying mass of the $\Delta$. 

In this work, we will also discuss the spin effects due to the triangle singularity as studied in Ref.\cite{Wang:2022wdm}. Here we shall study the SDMEs of the $\Delta$, which will be calculated in the quasi three-body decay process with taking the $M_\Delta$ at some fixed values and using the formula presented above. We shall consider the helicity states of the $\Delta$ in the c.m. frame of the $\gamma\Delta$ system. The spin density matrix element $\rho_{\lambda\lambda'}$ of the $\Delta$ as a function of the $\gamma\Delta$ invariant
mass in the $\gamma\Delta$ rest frame is defined as:

\begin{eqnarray}
\large{\rho}_{\smsm{\lambda\lambda'}}(m_{\gamma\Delta}) = \frac{\int {\rm d}
\Omega_{\bar p} {\rm d} \Omega_{\gamma}^* \sum\limits_{\text{spin}} {}'\CM^T_{\lambda}
\CM^{T*}_{\lambda'}}{\int{\rm d} \Omega_{\bar p} {\rm d}
\Omega_{\gamma}^*\sum\limits_{\text{spin}}|\CM^T|^2},\label{sdme}
\end{eqnarray}
where $\sum '$ represents the summing of all the spins apart from the $\Delta$'s, and $\lambda$ and $\lambda'$ are the helicities of the final $\Delta$. 

In this work, we will concentrate on the observable $P_\Delta$ defined as \begin{equation}
    P_\Delta = \frac{\rho_{11}-\rho_{33}}{\rho_{11}+\rho_{33}},\label{pdelta}
\end{equation}
where $\rho_{11}$ and $\rho_{33}$ are the diagonal SDMEs of the $\Delta$ and corresponding to the probability of finding the $\Delta$ in the helicity $\frac{1}{2}$ and $\frac{3}{2}$, respectively. Therefore, the $P_\Delta$ describes the asymmetry of the probabilities of the $\Delta$ having the helicities $\frac{1}{2}$ and $\frac{3}{2}$. Here we want to study the $M_{\gamma\Delta}$ dependence of the $P_\Delta$, so the angular dependence has been integrated(see Eq.\ref{sdme}). According to the definition, the value of the $P_\Delta$ can vary from $-$1 to 1. If TS mechanism dominates this reaction, we expect the $P_\Delta$ should approach 1 near TS. The $\rho^\Delta_{33}$ can be extracted from the angular distribution of its decay products, i.e. $\pi$ or $N$, in its rest frame through\cite{Thomas:1973uh}
\begin{equation}
    W(\cos\theta) = \frac{1}{4} \left[ \left( 1+4\rho_{33} \right) + \left( 3-12\rho_{33} \right) \cos^2\theta \right],
\end{equation}
and the $\rho_{11}$ can be deduced from the relation $\rho_{11}+\rho_{33}=\frac{1}{2}$.

\section{RESULTS AND DISCUSSION}
\label{results}

In this section, we shall study the TS mechanism in the reaction $J/\psi \to \gamma\bar{p}\Delta$ and discuss its effects on both the invariant mass spectrums of final particles and the $P_\Delta$.

\begin{figure}[htbp]
    \begin{center}
        \includegraphics[scale=0.48]{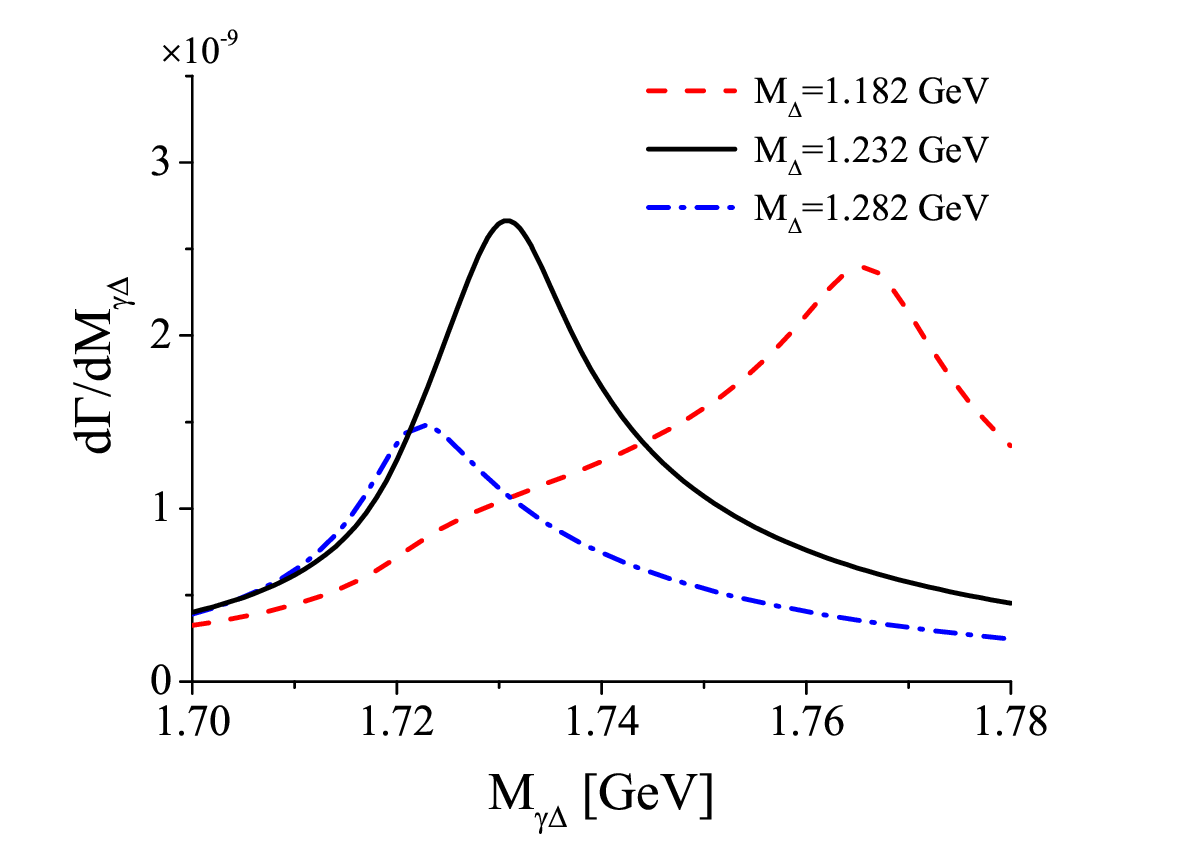}
        \caption{The distribution of the differential decay width versus the invariant mass $M_{\gamma \Delta}$ in $J/\psi \to \gamma \bar{p} \Delta$. The red dashed, black solid and blue dot-dashed lines denote the results with taking $M_{\gamma\Delta}$ = 1.182, 1.232 and 1.282 GeV, respectively.}
        \label{MS}
    \end{center}
\end{figure}

With using the package LoopTools\cite{Hahn:2000jm}, the loop integral in Eq.\eqref{Amplitude3} can be evaluated numerically. Through Eq.\eqref{Gamma3}, we can obtain the distribution of the differential decay width versus the invariant mass $M_{\gamma\Delta}$ by taking $M_\Delta=1.182$, $1.232$ and $1.282$ GeV individually. The corresponding results are depicted in Fig.\ref{MS}. As can be seen in the figure, the position of the peak caused by triangle singularity depends on the adopted mass of the $\Delta$. Therefore, by selecting the events in different region of the $M_{\pi N}$, the peak position in the invariant mass spectrum will change if TS mechanism indeed plays an important role here. As discussed in Ref.\cite{Bayar:2016ftu}, the moving peak observed here is mainly attributed to the reason that the position of TS is determined by kinematic conditions and dependent on the invariant mass of the external particles of the triangle loop. Following the method in Ref.\cite{Bayar:2016ftu}, by adopting the value of $M_\Delta$ from 1.081 to 1.286 GeV, the position of the TS in $M_{\gamma \Delta}$ can vary from 1.721 to 2.159 GeV. In fact,
there are two kinds of singularities which are relevant here\cite{Bayar:2016ftu}.
One is the normal two-body threshold cusp (TBTC), and
the other is the TS. In the case of $M_\Delta=1.182$ GeV(the red dashed line in \ref{MS}), the small bump around 1.73 GeV is caused by the TBTC. While, in other cases there is only one peak structure since the TS and TBTC are close to each other and their effects are overlapped. Here it is also worth noting that the width of the structure is rather narrow($\sim$20 MeV), which is mainly ascribed to the narrow width of the intermediate states in the loop. The feature of the moving peak and the rather narrow width of the peak structure caused by the TS mechanism therefore offer the clues for identifying the TS mechanism in experiment.

\begin{figure}[htbp]
    \begin{center}
        \includegraphics[scale=0.48]{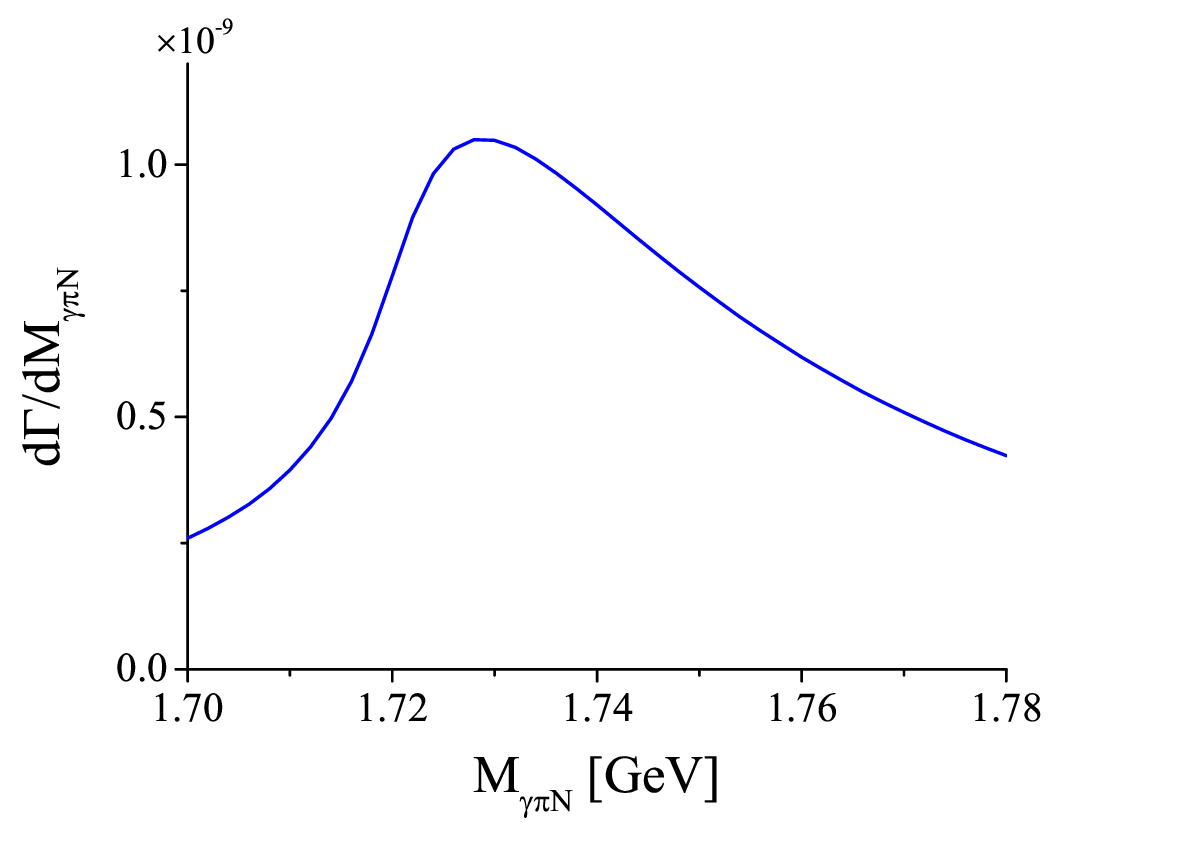}
        \caption{The distribution of the differential decay width versus the invariant mass 
$M_{\gamma\pi N}$ in $J/\psi \to \gamma \bar{p} \pi N$. }
        \label{MS4body}
    \end{center}
\end{figure}

Since the $\Delta$ is unstable and has a relatively large width, it is also necessary to further discuss the effects of its finite width on the invariant spectrum.  Based on the differential mass distribution formula in Eq.\eqref{4body}, we present the mass distribution as a function of $M_{\gamma\pi N}$ in Fig.\ref{MS4body} with considering the finite width effect explicitly. It can be found that with including the width effects of the $\Delta$ the peak structure become wider due to an average of the effects of the moving TS. While, even in this case the width of the structure is only about 30 MeV, which is significantly smaller than the width of the $N^*$ or $\Delta^*$ in this energy region and makes it distinguishable from ordinary resonance contributions. We can also calculate the decay branching ratio of $J/\psi \to \gamma\bar{p}\Delta$ using Eq.\eqref{Gamma3} with adopting $m_\Delta=1.232$ GeV, and we obtain
\begin{equation}
    \text{Br} \left( J/\psi \to \gamma\bar{p}\Delta \right) = 1.506 \times 10^{-6}.
\end{equation}
When futher considering the finite width of the $\Delta$ with taking $\Gamma_\Delta=0.117$ GeV, the decay branching ratio can be obtained through Eq.(\ref{4body}), then we get
\begin{equation}
    \text{Br} \left( J/\psi \to \gamma\bar{p}\Delta(\to \pi N) \right) = 1.058 \times 10^{-6}.
\end{equation} The production rate of this decay is within the measurable range at BESIII and also suitable to be explored at the Super Tau-Charm Facility.

\begin{figure}[htbp]
    \begin{center}
        \includegraphics[scale=0.48]{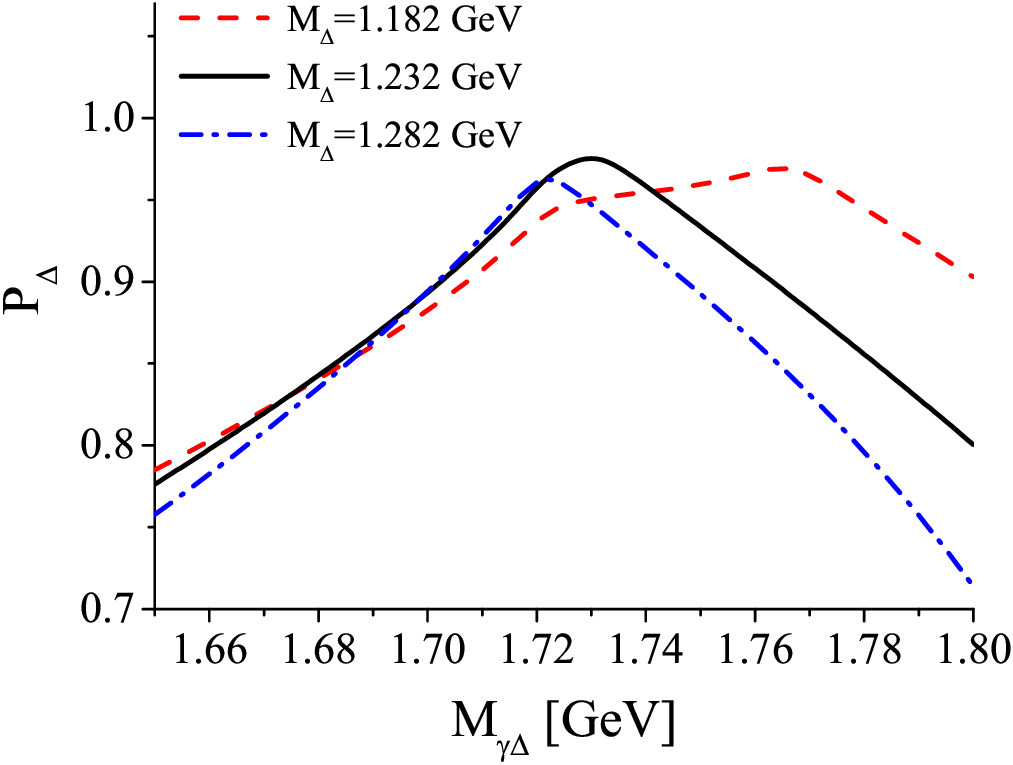}
        \caption{The SDME asymmetry $P_\Delta$ with regards to $M_{\gamma\Delta}$ for final $\Delta$ in reaction $J/\psi \to \gamma\bar{p}\Delta$.}
        \label{Psigma}
    \end{center}
\end{figure}

Next, let's focus on the spin effects induced by the TS mechanism on the $\Delta$. According to the Coleman-Norton theorem\cite{Coleman:1965xm}, TS occurs when the triangle loop process depicted in Fig.\ref{Feynman3} takes place in a classical manner. Specifically, in the rest frame of the $\gamma\Delta$ system, if the internal $\omega$, $\pi$ and $p$ are on-shell simultaneously, their three-momenta are collinear, and the $\pi$ moves in the same direction as the proton and can catch up with it to fuse to the $\Delta$, then the TS develops. Therefore, at TS the final $\Delta$ is predominantly produced by the intermediate $\pi$ and proton moving in the same direction as the final $\Delta$ in the $\gamma\Delta$ rest frame. In such a special condition, the $\Delta$ should be exclusively produced with helicity $\pm\frac{1}{2}$. To understand this result, it is helpful to consider the $\pi p$ elastic scattering in $s$-channel in the center of mass frame. In this process, even if the spin of the initial nucleon is unpolarized, the spin of the intermediate resonance is necessarily aligned when the spin of the intermediate resonance is larger than $\frac{1}{2}$\footnote{In the center of mass frame, if we take z axis along the direction of the momentum of the initial proton, the magnetic quantum number of the z component of orbital angular momentum has to be zero due to the fact that the momenta of the $\pi$ and $p$ are along z axis. Therefore, by taking the spin quantization axis along z axis, the spin projection along z axis of the intermediate state can only be $\pm \frac{1}{2}$ due to angular momentum conservation along z axis. For resonances with spin larger than $\frac{1}{2}$, it means that its spin is aligned.}. In the $J/\psi\to \gamma \bar p \Delta$ decay, since helicity is invariant under a boost from the $\Delta$ rest frame to the $\gamma \Delta$ rest frame, the above arguments also hold in the $\gamma \Delta$ rest frame. On the other hand, when the special kinematic conditions are not satisfied, i.e. departing the postion of the TS, the helicity of the $\Delta$ will not necessarily be $\pm \frac{1}{2}$ anymore. These expectations can be verified by a numerical calculation of the $P_\Delta$ defined above. In Fig.\ref{Psigma}, we show the $P_\Delta$ versus the $M_{\gamma\Delta}$ with taking the mass of the $\Delta$ as 1.182, 1.232 and 1.282 GeV, respectively. As can be seen from the figures, the $P_\Delta$ peaks appear at the corresponding TS positions in accordance with the expectations using the various $\Delta$ mass. Here, we want to note that such a $M_{\gamma\Delta}$ dependence is quite distinct from the expectation of a simple resonance model, since in resonance model the $M_{\gamma\Delta}$ dependence mainly comes from the denominator of the resonance propagator and should be canceled in calculating the ratio in Eq.\eqref{pdelta}. Therefore, the spin observable $P_\Delta$ can be used to verify whether the structure in invariant mass spectrum is caused by TS or a resonance. It is also interesting to notice that in the $M_\Delta=1.182$ GeV case(red dashed line in Fig.\ref{Psigma}) there is a small bump at about $M_{\gamma \Delta}=1.72$ GeV, corresponding to the $p\omega$ threshold, in the $P_\Delta$ distribution. As explained in Ref.\cite{Wang:2022wdm}, at $p\omega$ threshold the production of the $\Delta$ with the helicities $\pm\frac{1}{2}$ is also enhanced due to the kinematic condition. For the other cases, there is no such a structure due to the closeness of the $p\omega$ threshold and the TS. When considering $\Delta$ decay, we expect the peak structure of the $P_\Delta$ should still exist but with a larger width. However, by selecting the events in different $M_{\pi N}$ regions the phenomena discussed above should be observed in experiments.

Finally, when taking into account the $\Delta$ decay, the decay process $J/\psi \to \gamma\bar{p}\Delta(\to\pi^0 p/\pi^+ n)$ through the TS mechanism involves the $\pi^0 p\to \pi^0 p$ or $\pi^0 p\to\pi^+ n$ scattering as a subprocess. According to Schmid theorem\cite{Schmid:1967ojm}, in the  $\pi^0 p\to \pi^0 p$ case the contribution of the triangle loop diagram may be negligible compared to the corresponding tree level diagram. However, Ref.\cite{Debastiani:2018xoi} demonstrates that Schmid theorem holds strictly only in the limit $\Gamma_{\omega} \to 0$. Furthermore, by making a cut of the invariant mass $M_{\gamma\pi}$ in the final states it can also reduce the contribution of the tree diagram\cite{Huang:2020kxf}. In practice, it can also avoid the effects due to the Schmid theorem in this decay by choosing $\pi^+ n$ as the final state in experiment. Therefore, we expect the main features of the TS mechanism predicted in this work should still be observable after considering the Schmid theorem.

\section{SUMMARY}
\label{summary}

In this work, we investigate the triangle singularity developed in the $J/\psi \to \gamma\bar{p}\Delta$ process, where $\omega$, $\pi$ and $p$ compose the internal triangle loop. According to our results, the TS mechanism may induce a structure with a width of $0.02\sim 0.03$ GeV in the $\gamma\Delta$ invariant mass spectrum. We find the position of the TS is dependent on the $M_\Delta$ or the invariant mass of the final $\pi N$. By adopting the value of $M_\Delta$ ranging from 1.081 to 1.286 GeV, the position of the TS in $M_{\gamma \Delta}$ can vary from 1.721 to 2.159 GeV. Therefore, by performing a cut of the invariant mass of the final $\pi N$ the TS and the corresponding peak in the $M_{\gamma\Delta}$ distribution should be shifted accordingly. If the TS mechanism indeed plays an important role, we also expect that the spin observable $P_\Delta$ should take a relatively large value and have a peak versus the invariant mass $M_{\gamma\Delta}$ near the TS. The predicted decay branching ratio for this process is $\text{Br} \left( J/\psi \to \gamma\bar{p}\Delta(\to \pi N) \right) = 1.058 \times 10^{-6}$, which should be accessible at BESIII and future super Tau-Charm factory.

\begin{acknowledgements}

We acknowledge the support from the National Natural Science Foundation of China under Grants No.U1832160, the Natural Science Foundation of Shaanxi Province under Grant No.2019JM-025, and the Fundamental Research Funds for the Central Universities.
\end{acknowledgements}

\end{document}